\documentclass[sigconf]{acmart}
\usepackage{csquotes}
\usepackage{booktabs}
\usepackage{wrapfig}
\usepackage[T1]{fontenc}
\usepackage{enumitem}
\usepackage{array,multirow,graphicx}
\usepackage{amsmath}
\usepackage{amsfonts}
\usepackage{xcolor}
\usepackage{csquotes} 
\usepackage{tabularx}
\usepackage{makecell}
\usepackage{pbox}
\usepackage{subfigure}
\usepackage[hang,flushmargin]{footmisc}
\usepackage{flushend}
\usepackage{multicol, lipsum}
\usepackage[T1]{fontenc}
\usepackage{xspace}

\newcommand{\greatawakening}{{{\selectfont /v/GreatAwakening}}\xspace}
\newcommand{\qrv}{{{\selectfont /v/QRVoat}}\xspace}
\newcommand{\sqpatriots}{{{\selectfont /s/QPatriots}}\xspace}
\newcommand{\sga}{{{\selectfont /s/GreatAwakening}}\xspace}
\newcommand{\vnews}{{{\selectfont /v/news}}\xspace}
\newcommand{\snews}{{{\selectfont /s/News}}\xspace}
\newcommand{\sqstorm}{{{\selectfont /s/QStorm}}\xspace}
\newcommand{\rga}{{{\selectfont /r/greatawakening}}\xspace}
\newcommand{\rbeatingwomen}{{{\selectfont /r/beatingwomen}}\xspace}
\newcommand{\rcbts}{{{\selectfont /r/CBTS\_Stream}}\xspace}
\newcommand{\rtga}{{{\selectfont /r/The\_GreatAwakening}}\xspace}
\newcommand{\descr}[1]{\smallskip\noindent\textbf{#1}}

\definecolor{darkgreen}{RGB}{0, 100, 0}
\definecolor{linkcol}{RGB}{0,70,25}
\definecolor{citecol}{RGB}{0,70,25}
\definecolor{urlcol}{RGB}{0,70,25}
\definecolor{celestialblue}{RGB}{0.29, 0.59, 0.82}
\definecolor{darkblue}{rgb}{0,0,0.5}

\newlist{mycompactitem}{itemize}{1}
\setlist[mycompactitem,1]{
  label=--,         
  left=1pt,          
  labelindent=0em,  
  itemindent=0em,    
  parsep=0pt,         
}

\AtBeginDocument{%
  \providecommand\BibTeX{{%
    \normalfont B\kern-0.5em{\scshape i\kern-0.25em b}\kern-0.8em\TeX}}}

\copyrightyear{2024}
\acmYear{2024}
\setcopyright{rightsretained}
\acmConference[Websci '24]{ACM Web Science Conference}{May 21--24, 2024}{Stuttgart, Germany}
\acmBooktitle{ACM Web Science Conference (Websci '24), May 21--24, 2024, Stuttgart, Germany}\acmDOI{10.1145/3614419.3644021}
\acmISBN{979-8-4007-0334-8/24/05}

\begin{document}

\title{Waiting for Q: An Exploration of QAnon Users' Online Migration to Poal in the Wake of Voat's Demise}


\author{Antonis Papasavva}
\affiliation{%
  \institution{University College London}
  \city{London}
  \country{United Kingdom}}
\email{antonis.papasavva@ucl.ac.uk}

\author{Enrico Mariconti}
\affiliation{%
  \institution{University College London}
  \city{London}
  \country{United Kingdom}
}
\email{e.mariconti@ucl.ac.uk}


\begin{abstract}
Online communities are groups of people who interact primarily via the Internet, often sharing common interests. 
Some of these groups, particularly supporters of \emph{Q} who created the far-right conspiracy theory known as \emph{QAnon}, are highly toxic and controversial. 
These communities are often banned from various mainstream online social networks due to their controversy. 
This study examines the deplatforming and subsequent migrations of QAnon adherents, following a two-step process.
We analyze Reddit data, finding that users opt for Voat as an alternative following the Reddit bans, particularly influenced by Q's postings on 4chan. 
Subsequently, upon Voat's shutdown announcement, we observe users recommending \emph{Poal}. 
Among several insights, we compare the effects of abrupt permanent bans and announced shutdowns on the migration patterns of these conspiracists. 
Specifically, we find that almost half of Poal's active users are Voat migrants who registered after the shutdown was announced. 
This contradicts the patterns observed after Reddit bans, suggesting that advance warning can facilitate more coordinated migrations. 
Lastly, our research uncovers evidence of discussions and planning related to the January 6th, 2021, attack on the US Capitol, which emerged shortly after Voat's shutdown, predominantly on Poal.
This underscores the continued activity of the conspiracy, albeit at a diminished scale due to various bans and a shutdown, while also exposing Poal as a platform that hosts dangerous individuals.

\end{abstract}


\maketitle

\section{Introduction}
Online social networks have rules, terms, and conditions that users often do not abide by, which, in turn, requires the content moderators to take action. 
Content moderation is the organized governance, detection, assessment, and intervention taken on user-created content or behavior that stands against a platform's policies~\cite{gillespie2020expanding}.
A social network known to have faced challenges related to hateful speech, self-harm promotion, and misinformation is Reddit~\cite{reddit2018hate}, as it allows users to create and moderate their own communities, leading to a wide range of topics, including those criticized for promoting controversial ideologies. 
As part of its efforts to address problematic communities, Reddit began permanently banning them (\emph{deplatforming}), with the first being \rbeatingwomen in 2014~\cite{reddit2014beatingwomen}.

However, banning practices does not eliminate overall online toxicity as banned communities frequently migrate to other platforms and, in some cases, become more toxic~\cite{ribeiro2021migrations}.
Voat was a Reddit-like news aggregator launched in April 2014 that encouraged free speech and gained attention as a Reddit alternative after the bans of controversial subreddits~\cite{mekacher2022can}, especially QAnon-focused ones~\cite{papasavva2021qoincidence}. 

QAnon is a conspiracy theory that originated on 4chan's Politically Incorrect board on October 28, 2017~\cite{papasavva2022gospel}. 
The posts that ``Q,'' the creator of this conspiracy, made on 4chan and 8chan are referred to as ``Q drops.'' 
In these cryptic messages, Q explains, among other things, how a network of blood-thirsty pedophile politicians, celebrities, and influential people are part of the deep state: an undercover organization that controls and runs governments worldwide~\cite{sky2020whatisq}. 
At the same time, allegedly, Donald Trump struggles to uncover the deep state and bring its members to justice.

Voat struggled financially and had to shut down on December 25, 2020, announcing it on December 21.\footnote{\url{https://searchvoat.co/v/announcements/4169936}}
With less than four days to organize their migration, many users expressed their intent to move to \emph{Poal.co}, which quickly emerged as a popular alternative for Voat users. 
Poal is a Reddit-like social network with a ``\emph{say what you want}'' slogan, where discussions are divided into ``subs,'' akin to subreddits. 
In this work, we investigate the unique scenario of Voat's planned shutdown and the subsequent organized migration to Poal, offering insights into online community migration dynamics. 

\descr{Motivation.}
Studying user migration patterns and behavior during moderation efforts or shutdowns is crucial in understanding the dynamics of online communities and the impact of such events on alternative platforms. 
There is limited understanding of the specific factors influencing user migration when faced with a shutdown.

Voat's shutdown announcement provides a unique opportunity to investigate the above research gap. 
By examining how users organized their migration and exploring the subsequent activity on alternatives, we gain insights into the strategies and preferences of users navigating a changing online landscape.
We focus on QAnon communities since they were the most active on Voat~\cite{mekacher2022can}, and we aim to compare the migration patterns of Reddit to Voat and Voat to Poal as we believe such analysis offers valuable insights into the dynamics of user migration. 
Another reason we choose to study the migration patterns between Reddit, Voat, and Poal is due to the visual and functional similarities of these social networks and the reports~\cite{papasavva2021qoincidence, mekacher2022can} of Voat gaining popularity once Reddit decisively banned all QAnon subreddits.

\noindent Our contributions are as follows:
\begin{mycompactitem}[noitemsep,topsep=0pt]
\item \textbf{Unique Study of Organized Online Migration}: This research is the first to investigate the distinctions between the sudden Reddit-to-Voat migration and the premeditated Voat shutdown, where alternatives were suggested. Notably, Poal emerged as a popular alternative. Overall, we study a two-hop migration between three different platforms.
\item \textbf{Expose Poal as a Platform Hosting Dangerous Individuals}: We provide a novel analysis of Poal, an alternative social network similar to Reddit and Voat. This platform's stance on ``free speech'' makes it attractive to banned and fringe communities. Our findings reveal that QAnon adherents on Poal, many of whom migrated from Voat, began planning the January 6th, 2021, US Capitol attack shortly after Voat's shutdown. This highlights the significance of studying fringe communities that lack comprehensive understanding.
\item \textbf{Broadening Previous Findings}: Our study reaffirms previous findings as we observe a decrease in the user base of the QAnon conspiracy theory with each ban, consistent with Ribeiro et al.~\cite{ribeiro2021migrations}. However, our research offers a broader examination of QAnon migration aspects. We explore user discussions about alternatives, their opinions, and the impact of unexpected bans versus announced ones. While only half of Voat's QAnon user base migrated to Poal, this migration appears more organized than previous QAnon migrations, highlighting user behavior patterns during migration, especially when premeditated.
\end{mycompactitem}

\section{Related Work}\label{sec:related-work}
We set to discuss related work on deplatforming and online user migration, along with works that focus on the QAnon conspiracy.

\descr{Deplatforming.}
One of the first works on Reddit migrations is the one of Chandrasekharan et al.~\cite{chandrasekharan2017you}.
This work analyzes the ban of r/fatpeoplehate and r/CoonTown, finding that these bans successfully lowered Reddit's toxicity, as most users stopped using Reddit, and the ones that stayed drastically reduced hateful messages. 
A similar study~\cite{newell2016user} focused on the 2015 Reddit bans to analyze changes in user behavior, observing small migrations to various platforms, highlighting that not many users left the platform.

Engel et al.~\cite{engel2022characterizing} study the early QAnon supporters on Reddit.
After the 2018 QAnon subreddit bans, the users that remained active on Reddit mostly engaged with other sympathetic towards the conspiracy subreddits.
This study demonstrates that many users cling to the same platform, even after their community is banned. 
Similarly, in our study, we find that when \rcbts was banned, the activity of \rga, another QAnon-focused subreddit, peaked at levels higher than that of \rcbts.

Previous works~\cite{trujillo2022make,jhaver2021evaluating} explain that although bans prove to be sufficient to eliminate specific discussions and users from one platform, those users tend to migrate to other more lenient platforms.
More specifically, Ribeiro et al.~\cite{ribeiro2021migrations} collect data from Reddit's r/The\_Donald and r/Incels that reportedly migrated to their standalone platforms.
The authors show that Reddit bans cause only fractions of the banned userbase to migrate to other platforms, leading to reduced activity, while, the r/The\_Donald community demonstrated higher toxicity on their new platform, similar to the findings of Ali et al.~\cite{ali2021understanding}, which focuses on Gab users that were suspended from Twitter and Reddit.
Similarly, Habib et al.~\cite{habib2022proactive} explain that Reddit administrators struggle to prevent or contain toxic activity and facilitate early interventions, which is more appropriate than banning or quarantining, aligned with the findings of Gilbert~\cite{gilbert2013widespread}.
Chandrasekharan et al.~\cite{chandrasekharan2022quarantined} collect $85M$ posts from Reddit's r/TheRedPill and r/The\_Donald subreddits to examine the quarantine’s effects on them, finding that quarantining subreddits makes it difficult to recruit new members, but the existing userbase remained as controversial as before.
Finally, Russo et al.~\cite{russo2023understanding} study the post-ban migration of r/The\_Donald and r/fatpeoplehate subreddits, finding that toxic users are more likely to migrate to other platforms after a ban and are likely to remain in the previous platform at the same time, remaining active in both platforms.

\descr{QAnon.}
One of the first studies that detected QAnon discussions online was the one of Darwish~\cite{darwish2018kavanaugh}, who analyzed tweets related to US Supreme Court judge Brett Kavanaugh, finding QAnon-related hashtags amongst the top 6 popular of their dataset.
McQuillan et al.~\cite{mcquillan2020cultural} collected $81M$ tweets related to COVID-19 and reported that the QAnon movement has grown throughout the pandemic and has reached more mainstream groups.

Our prior study on Voat was the first quantitative investigation into the QAnon community's online discourse and toxicity~\cite{papasavva2021qoincidence}. 
It showed that the audience of \greatawakening mainly consumed content from a few creators, aligning with findings by Priniski et al.~\cite{priniski2021rise}.
Our analysis identified QAnon-related keywords and discussions, focusing on US politics, Donald Trump, and conspiracy theories.
Subsequent research aimed to understand the dissemination and discussion of content attributed to the central figure of the conspiracy, 'Q'~\cite{papasavva2022gospel}, finding that Q's posts are incoherent and not toxic or threatening, but the interpretations of adherents imbue them with dangerous potential. Additionally, our findings suggest Q may not be a singular individual, as indicated by OrphAnalytics\cite{OrphAnalytics2020}. 
Lastly, despite Reddit's ban on popular QAnon subreddits, discussions around Q's posts persist on the platform, albeit less frequently.

Phadke et al.~\cite{phadke2021characterizing} collect Q drops and posts from 12 QAnon-related subreddits to study, among other things, the social imaginaries established by Q.
The authors explain that QAnon followers express discomfort when their beliefs are challenged.
Sharma et al.~\cite{sharma2022characterizing} study the online engagement with disinformation and conspiracies during the 2020 US Presidential Elections on Twitter.
The authors explain that many accounts (both right-leaning and left-leaning) engaged with QAnon tweets, which suggests that conspiratorial and extreme narratives can potentially worsen ideological divisions.
Zihiri et al.~\cite{zihiri2022qanon} analyzed over $3.5M$ messages shared on $94$ Telegram communities, finding that QAnon is a distinct political movement that interacts with existing extremist groups, especially during significant political events.
The authors observe that QAnon followers often cite mainstream information sources that often align with conservative views.
Sipka et al.~\cite{sipka2022comparing} analyze Twitter, Parler, and Gab data, finding that Parler had the highest volume of posts, Twitter had more unique users posting with QAnon hashtags, and Gab had the highest proportion of posts with hate words.

Hanley et al.~\cite{hanley2022no} investigated QAnon website relationships, finding that many online platforms have taken steps to curb the spread of the QAnon conspiracy theory, but it continues to be hyperlinked by mainstream sources like Twitter, YouTube, and major news outlets.
Also, QAnon websites are well connected, making it easy for users to discover additional websites that promote conspiracy.
B{\"a}r et al.~\cite{bar2023finding} analyze over $600K$ Parler user profiles and the content they created, finding that almost $35K$ users openly support QAnon, having on average more followers,
followees, and posts, than the non-supporters.
Hoseini et al.~\cite{hoseini2023globalization} collect $4.4M$ messages shared in $161$ QAnon groups on Telegram to study the globalization of the conspiracy on the platform.
The authors find that QAnon content and the number of active QAnon groups on Telegram has been increasing.
Finally, Imperati et al.~\cite{imperati2023conspiracy} attempt to quantify how conspiracy theories raise funds by exploiting Telegram. 
The authors find, among other things, that conspiracy-focused channels actively seek to profit from their subscribers, resulting in raised funds of \$90 million by arranging crowdfunding campaigns.

\begin{table}
\centering
\small
\smallskip
\begin{tabular}{l | l r r r}
\toprule
& \textbf{Community} & \textbf{\# S} &\textbf{\# C} & \textbf{Date Span} \\
\midrule
\parbox[t]{2mm}{\multirow{2}{*}{\rotatebox[origin=c]{90}{Reddit}}}
& \rcbts & 30,176 & 267,744 & 20 Dec 17 - 28 Feb 18 \\
& \rga & 79,952 & 926,676 &  9 Jan 18 - 19 Aug 18 \\ 
\\[-0.8em]
\hline
\parbox[t]{2mm}{\multirow{3}{*}{\rotatebox[origin=c]{90}{Voat}}}
& \vnews & 176,948 & 1,397,955 & 1 Oct 16 - 25 Dec 20 \\
& \greatawakening & 100,699 & 982,702 & 9 Jan 18 - 25 Dec 20 \\
& \qrv & 185,929 & 2,225,702 & 22 Sep 18 - 25 Dec 20 \\
\hline
\parbox[t]{2mm}{\multirow{4}{*}{\rotatebox[origin=c]{90}{Poal}}}
& \snews & 26,919 & 74,431 & 29 Mar 18- 6 Sep 21 \\
& \sga & 1,889 & 2,032 & 13 Jul 19 - 6 Sep 21 \\
& \sqpatriots & 5,930 & 22,167 & 09 Feb 21 - 6 Sep 21 \\
& \sqstorm & 19,966 & 109,358 & 12 Jul 19 - 6 Sep 21 \\
\toprule
\end{tabular}
\caption{Number of submissions (S) and comments (C) collected from Reddit, Voat, and Poal.}
\label{tbl:datasets}
\vspace{-0.5cm}
\end{table}

\section{Datasets}\label{sec:datasets}
Table~\ref{tbl:datasets} presents dataset dates and post counts by community.

\descr{Reddit.}
We collect Reddit data from Pushift~\cite{baumgartner2020pushshift}, by searching for the three popular QAnon-focused subreddits: \rga, \rcbts, and \rtga~\cite{redditTOvoat}.
Since the Pushift dataset for \rtga is incomplete; only $1.4K$ submissions without any comment data, we use Pushift data from \rga and \rcbts, only. 
We intentionally omit the collection of a baseline subreddit, such as `News,' as our primary objective is to investigate the impact of Reddit bans on Voat and Poal, with a specific focus on QAnon-related subverses and general-discussion subverses within those two platforms. 
Analyzing the effect of Reddit bans within Reddit is beyond the scope of our study.

\descr{Voat.}
Our Voat dataset includes data from \greatawakening and \qrv, the two most popular QAnon subverses on Voat, and \vnews, which we use as a baseline subverse to compare the overall Voat activity.
We collect the entire history of these subverses from the public Voat dataset published by Mekacher and Papasavva~\cite{mekacher2022can}.
We consider \vnews data created between October 2017 and December 2020, disregarding activity before the surface of QAnon.

\descr{Poal.}
We implement a custom crawler to collect Poal data following the methodology of Papasavva et al.~\cite{papasavva2021qoincidence}; a DOM-tree scraper using HTML requests and Beautiful Soup.
Our crawler operated between July 1, 2021, and September 7, 2021, and collected data in reverse order, starting from the most recent post of every subverse's first page.
To guarantee our dataset's completeness, our crawler followed a list of subverses marked for collection, going through the entire history of every subverse on the list in a loop.
After the list of subverses was exhausted, the crawler repeated this process constantly, accessing submissions it had already collected and looking for new comments.
To identify QAnon subverses on Poal, the lead author visited the full list of Poal subverses and checked every subverse for possible QAnon association.\footnote{\url{https://poal.co/subs}}
After careful inspection, we found tens of subverses devoted to QAnon but mostly inactive.
To this end, we focus on three active and popular Poal QAnon-focused subverses, namely \sga, \sqpatriots, and \sqstorm.
As a baseline, we collect \snews.

\subsection{Ethical Considerations}\label{sec:ethics}
\descr{Content Warning.}
This manuscript includes direct quotations of offensive language and slurs from the platforms analyzed to represent their linguistic landscape accurately. 
We disapprove of using such language and do not endorse or align with the beliefs associated with these slurs.

\descr{Data Collection.}
For our Reddit and Voat datasets, we use publicly available data published, discussed, and analyzed in previous works~\cite{mekacher2022can,papasavva2022gospel}.
Regarding Poal data collection, we followed the Terms and Conditions of the site and respected their operational data crawling services. 
The authors and crawler did not interact with online users in any way nor used a registered user-bot (simulated logged-in activity) to collect data. 

\descr{Username Matching.}
In Section~\ref{subsec:useractivity}, we perform username matching across Reddit, Voat, and Poal to examine user migrations between the three platforms.
To safeguard the anonymity of users, we encrypted the usernames using the SHA256 algorithm, keeping the original usernames confidential.
Similar approaches have been proposed and used in prior research~\cite{ribeiro2021migrations,rajadesingan2020quick}, which we follow in attempting to minimize the risk of user de-anonymization.

Our methodology involves exact matching on publicly available data, and at no point do we track or single out individual usernames.  
This approach is designed to respect reasonable privacy expectations and mitigate any concerns related to the infringement of user anonymity. 

\descr{Ethical Review.}
UCL's Department of Security and Crime Science ethics committee approved the collection of the Poal dataset and the use of the public Reddit and Voat datasets, along with the use of these datasets, to perform the analysis discussed in this paper.

\begin{table}
\centering
\smallskip
\begin{tabular}{l l r }
\toprule
\textbf{\#} & \textbf{Date} & \textbf{Event} \\
\midrule
1 & October 28, 2017 & Q posts their first Q drop \\
2 & March 14, 2018 & Reddit bans \rcbts \\
3 & September 12, 2018 & Reddit bans \rga \\
4 & December 25, 2020 & Voat shuts down \\
5 & January 6, 2021 & United States Capitol attack \\
\toprule
\end{tabular}
\caption{QAnon important dates.}
\label{tbl:important_dates}
\end{table}

\begin{figure*}[t]
\centering
\subfigure[Submissions]{\includegraphics[width=0.8\textwidth]{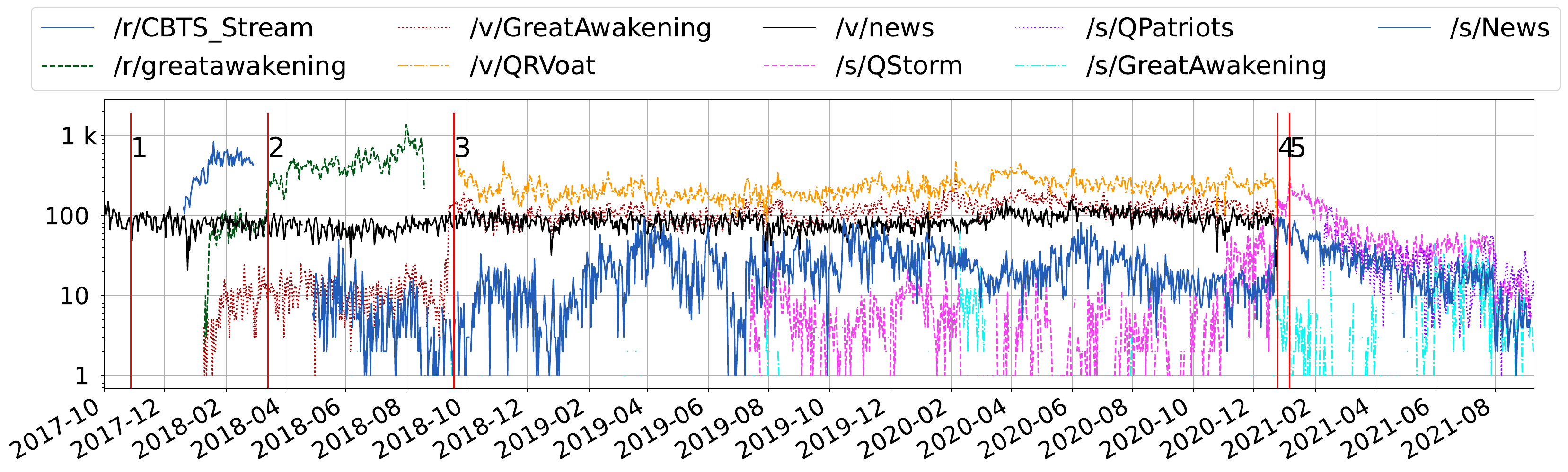}\label{fig:all_submissions}}
\subfigure[Comments]{\includegraphics[width=0.8\textwidth]{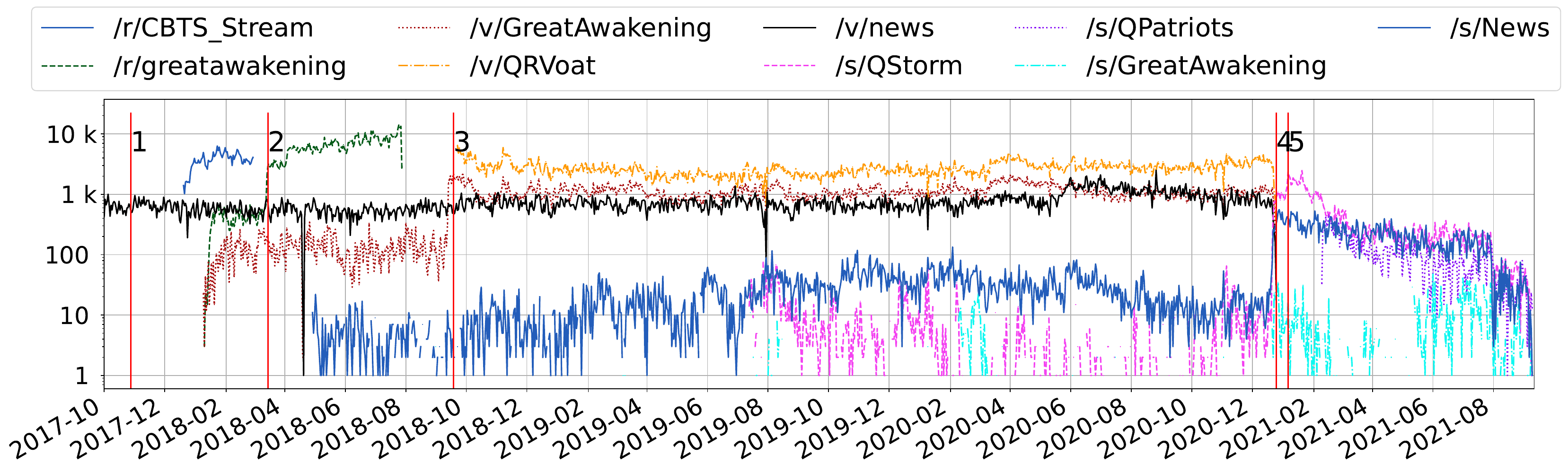}\label{fig:all_comments}}
\caption{Number of (a) submissions and (b) comments per day on Reddit, Voat, and Poal.}
\label{fig:all_comments_submissions}
\end{figure*}

\section{Longitudinal Analysis}\label{sec:analysis}
This section illustrates the evolution of adherents' activity longitudinally across three platforms.
We focus on how QAnon adherents organize their online migration after a shutdown.
To this end, we highlight five major events related to the conspiracy's evolution listed in Table~\ref{tbl:important_dates}, which we analyze throughout this paper.

\begin{figure*}[t]
\centering
\subfigure[Submitters]{\includegraphics[width=0.8\textwidth]{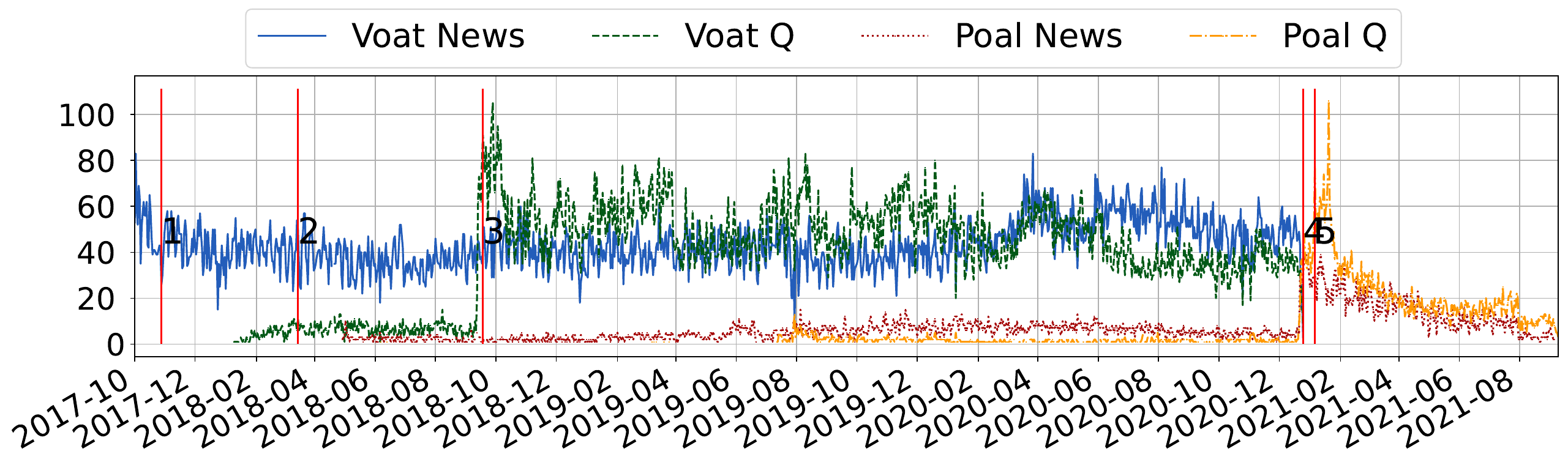}\label{fig:submitters}}
\subfigure[Commenters]{\includegraphics[width=0.8\textwidth]{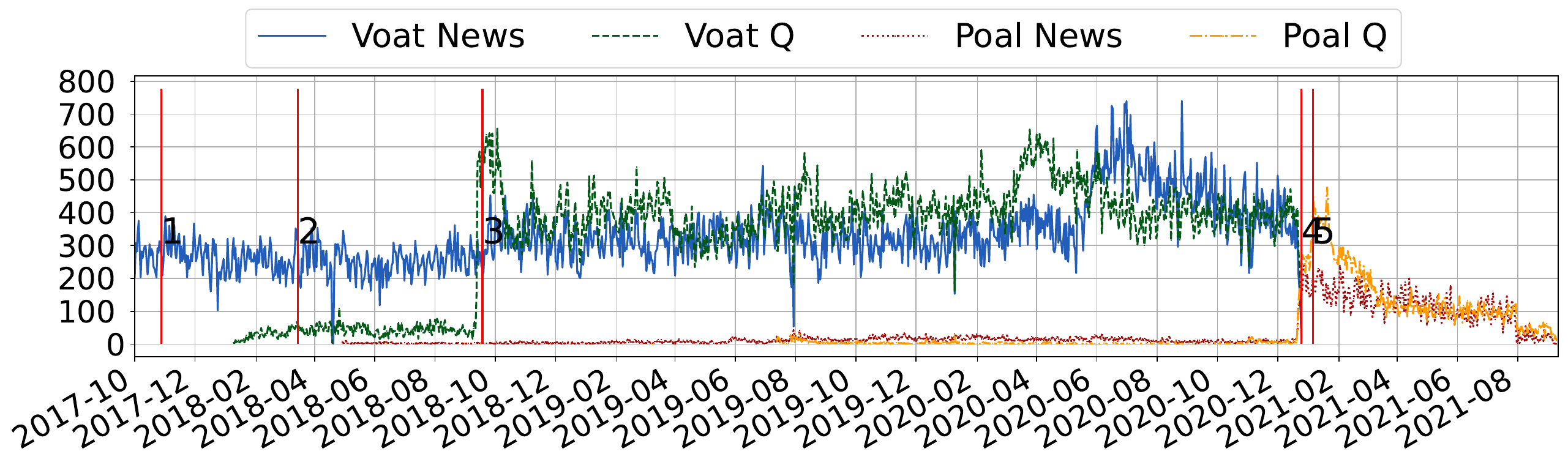}\label{fig:commenters}}
\caption{Number of (a) submitters and (b) commenters per day on Voat and Poal.}
\label{fig:all_commenters_submitters}
\end{figure*}

\subsection{Temporal Analysis}\label{sec:temporal}
First, we plot the number of submissions and comments on Reddit, Voat, and Poal in Figure~\ref{fig:all_comments_submissions}.
Note the log scale on the x-axis, and the red vertical lines marking the important dates shown in Table~\ref{tbl:important_dates}.
Figure~\ref{fig:all_submissions} plots the number of submissions on the three platforms.
QAnon discussions appear on Reddit on December 20, 2017, with the first submission on \rcbts.
The subreddit \rga appears on January 9, 2018, which is the same date that \greatawakening makes its appearance on Voat as well.

The \rcbts subreddit went silent on February 28, 2018, due to Reddit quarantining it before banning it on March 14, 2018, resulting in the activity of \rga increasing tenfold, likely due to \rcbts users using that subreddit for their discussions.
Meanwhile, on June 7, 2018, \sga posted its first-ever submission on Poal.
On August 19, 2018, Reddit quarantined \rga before banning it on September 12, 2018.
Immediately after \rga is banned, the activity on \greatawakening increases from approximately $90$ comments daily to more than $1K$, on average, daily after the ban (Figure ~\ref{fig:all_comments}).
Notably, ten days after the \rga ban, \qrv is created on Voat, posting more than $2.3K$ comments, on average, daily; more than double the pre-existing QAnon subverse on Voat.

Overall, the subreddits \rcbts and \rga, and the subverse \greatawakening needed some time before gaining popularity. 
On the contrary, \qrv quickly became the center of QAnon discussions on Voat from its inception. 
This immediate surge suggests coordinated action, prompting us to investigate its origin. 
After inspecting Voat's activity on that day, we discovered links to three different Q drops posted on 4chan on September 21, 2018, urging users to migrate to \qrv as an alternative platform for displaced users.\footnote{\url{https://qanon.pub/\#t1537581143},~\url{https://qanon.pub/\#t1537581556}, and~\url{https://qanon.pub/\#t1537576793}}
This revelation was surprising and concerning, revealing the swift and organized nature of the QAnon community across multiple platforms.
Furthermore, this incident underscores a broader trend of 4chan's influence beyond its platform. 
Notably, 4chan users have been known to orchestrate raids on YouTube videos, flooding them with hateful comments to bring them down~\cite{mariconti2019you}. 
This finding underscores the significant impact of 4chan, despite its fringe status, on various online platforms.

Turning to Poal, the subverse \sqstorm appeared on September 2, 2019, and gained popularity near the end of 2020, after Voat's co-founder announced its shutdown (event 4 in Table~\ref{tbl:important_dates}). 
After Voat shuts down, the activity on \sqstorm and \sga rises.
Lastly, \sqpatriots appears on Poal on February 9, 2021, and becomes the second most popular subverse on the platform after \sqstorm.
Similar to \qrv taking over \greatawakening on Voat, \sqstorm seems to take a second best place on Poal even though \sga existed since June 2018.
Regarding the baseline subverses: \vnews and \snews; we notice that the Reddit bans did not affect Voat's baseline, but that is not the case for \snews, which appears to be more active after Voat shuts down.

\descr{Key findings.}
On Reddit, users moved from \rcbts to \rga instead of moving to Voat, showing clear signs that users prefer to stay in a platform that would likely ban them eventually, rather than move to a different one.
The activity of \greatawakening increased by ten times after both Reddit bans.
More importantly, \qrv appeared long after and took over \greatawakening as the most active QAnon subverse on Voat, following Q's recommendation on 4chan.
Similarly, \sqpatriots appears after \sqstorm and \sga but is almost as popular as \sqstorm from its first week.
This activity shows that the creation of specific communities was coordinated. 

\subsection{User Activity}\label{subsec:useractivity}
Taking a closer look at user activity, we plot the unique number of submitters and commenters per day on Voat \vnews, and QAnon subverses, along with Poal \snews and QAnon subverses in Figure~\ref{fig:all_commenters_submitters}.
We omit Reddit data from this figure for better visualization.
Again, the vertical red lines indicate the key events from Table~\ref{tbl:important_dates}.

Examining QAnon subverses on Voat (\emph{Voat Q} in Figure\ref{fig:submitters}), we note that the range of daily submitters before September 11, 2018, is between 1 and 13.
Immediately after the last QAnon subreddit was banned on September 12, 2018, the average number of daily submitters on Voat QAnon subverses rose to $43.2$.
Notably, the baseline subverse \vnews receives popularity after the Reddit bans, but not for long.
More specifically, the average number of \vnews submitters was $42.7$ before the Reddit bans and rose to $43.2$ after the bans, indicating no significant change.
Turning to previous work~\cite{mekacher2022can}, it is unsurprising that \vnews receives virtually no additional activity since many fringe communities on Voat refrain from engaging with other communities.

We notice a peak in \vnews activity between May 2020 and October 2020.
Manual inspection of popular posts during that period indicates that this peak is due to new and old users discussing various news and events, including, among others, gun control in Canada~\cite{guncontrol}, the COVID lockdowns, Donald Trump denying links to Venezuela armed raid by US citizens~\cite{venezuela}, and the George Floyd BLM protests~\cite{george}.
Considering the worldwide lockdowns, observing a peak in user activity during that period is not unexpected. 

Once the Voat shutdown was announced on December 21, 2020, the numbers of unique submitters and commenters remained stable.
On the contrary, the activity of unique submitters and commenters per day on both QAnon and baseline subverses on \emph{Poal} peaked immediately on December 21, 2020.
Specifically, the average number of submitters on the Poal QAnon subverses and on \snews before the announcement was $4.8$ for both, and they peaked at $239.7$ and $25.3$, respectively, per day after the announcement.

\descr{Key findings.}
Following the final Reddit ban, Voat experienced a significant increase in daily submitters on its QAnon subverses, suggesting a migration of QAnon adherents to Voat as an alternative platform. 
Despite the anticipated increase in activity and discussions on Voat as its shutdown approached, there was no significant change in the number of unique submitters and commenters. 
In contrast, on Poal, both QAnon and baseline subverses experienced a peak in activity immediately after the Voat shutdown announcement, with a substantial increase in daily submitters on Poal QAnon subverses.

\begin{figure}[t]
\centering
\includegraphics[width=0.7\columnwidth]{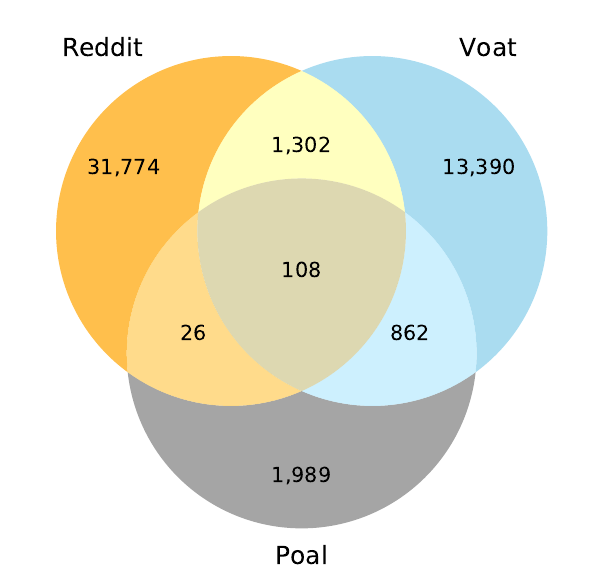}
\caption{Intersection of usernames on Reddit, Voat, and Poal QAnon communities.}
\label{fig:user_intersection}
\end{figure}

\descr{Username intersection.} 
We now compare the usernames across all three platforms to detect any exact matches. 
We hypothesize that many Voat users rushed to register a new account on Poal since the platform was mentioned more than 100 times in the Voat shutdown announcement post.\footnote{\url{https://archive.searchvoat.co/v/announcements/4169936}}

For this analysis, we plot the \emph{case insensitive}, \emph{identical} username intersection Venn diagram in Figure~\ref{fig:user_intersection}.
To protect the anonymity of users, we encrypt all usernames across all three platforms using the hash function \emph{SHA256}.\footnote{\url{https://docs.python.org/3/library/hashlib.html}}
This guarantees that we do not track raw usernames across platforms but rather demonstrate the existence of identical usernames across them.
For more details regarding our ethics considerations, please refer to Section~\ref{sec:ethics}.

$108$ usernames are identical across all three platforms.
Also, Reddit and Voat share $1.3K$ usernames ($9.7\%$ of Voat QAnon engaging users), while Voat and Poal share $862$ usernames ($43.3\%$ of Poal QAnon engaging users).
Although the percentage of username matches is not significant between Reddit and Voat ($9.7\%$), it is clear that Poal attracted a noteworthy amount of users from Voat ($43.3\%$).
This analysis indicates that about half of the users on Poal had an account on Voat.
Notably, our findings are aligned with previous work~\cite{ribeiro2021migrations} that shows how activity on new platforms or the appearance of newcomers decreases after a ban.
Looking at the face value of the numbers in Figure~\ref{fig:user_intersection}, if we consider that Reddit's QAnon user base was $31.7K$, then Voat only had $42.1\%$ of QAnon users after the Reddit bans.
After the Voat shutdown, Poal only managed to attract $14.8\%$ of users compared to its predecessors, Reddit and Voat.
This is because the community is more dispersed and smaller after every ban, or shutdown in this case. 
Nevertheless, the migration from Voat to Poal seems more organized than from Reddit to Voat. 

\descr{Key findings.}
There is a significant overlap of usernames between Voat and Poal, indicating migration of Voat users to Poal. 
The overlap of usernames between Reddit and Voat is less significant, suggesting a lower level of migration between these platforms. 
The analysis also highlights that approximately half of the users on Poal previously had an account on Voat. 
Overall, the findings demonstrate decreased QAnon communities with each ban or shutdown.

\section{Topic evolution}\label{sec:bigrams}
We now investigate how discussions on Reddit, Voat, and Poal changed through popular bigram analysis during the Reddit bans and the Voat shutdown announcement. 
For Reddit, we perform bigram analysis on two specific QAnon communities: \rcbts and \rga. 
The analysis for \rcbts covers data from five days before its quarantine (Feb 23 - Feb 28, 2018), while for \rga, we consider data from five days after \rcbts is banned (Mar 14 - Mar 19, 2018) and five days before \rga is quarantined (Aug 14 - Aug 19, 2018). 
For Voat, we consider data from five days before the Voat shutdown announcement (Dec 16 - Dec 20, 2020) and data from five days after the announcement (Dec 21 - Dec 25, 2020). 
In the case of Poal, we examine discussions from five days before the Voat shutdown announcement (Dec 16 - Dec 20) and extend our analysis for ten days after the announcement (Dec 21 - Dec 30). 
The extended analysis aims to capture discussions about Voat's shutdown and user activity on Poal after migrating there. 
We focus on the final five days of each community to explore discussions about alternative social networks as their communities near the end of their lifetimes. 

Initially, we explored various topic modeling methods, including BTM, LDA, and BERT, for this analysis; however, due to limited data during the short time frames, these approaches did not yield coherent results.
To this end, we employ bigram analysis and manually review posts containing specific bigrams.
Following a methodology inspired by Gauthier et al.~\cite{creswell2017designing}, we treated bigrams as thematic codes within the selected posts. 
Specifically, we used purposeful sampling to identify posts with specific bigrams. 
In our review process, we randomly selected a representative subset (half unless stated otherwise) of posts for examination, focusing on meaningful concepts and connections. 
This analysis involved discussions among both authors, drawing on our knowledge of the conspiracy theory, previous research, and contextual events.

\begin{table}[!tbp]
\centering
\footnotesize
{\begin{tabular}{l r | l r | l r }
\toprule
\multicolumn{2}{c}{CBTS before quarantine} &  \multicolumn{2}{|c}{GA after CBTS ban} & \multicolumn{2}{|c}{GA before quarantine} \\
\midrule
\textbf{Bigram} & \textbf{\emph{f}} & \textbf{Bigram} & \textbf{\emph{f}} & \textbf{Bigram} & \textbf{\emph{f}}  \\
\midrule
q post & 151  & deep state & 122 & q post & 65 \\
deep state & 143 & q post & 96 & red shoe & 46 \\
false flag & 81 & social media & 87 & q drop & 31 \\
god bless & 76 & bot action & 65 & security clearance & 27 \\
social media & 71 & action perfromed & 65 & wwg wga & 26 \\
red pill & 67 & cbts stream & 61 & anderson cooper & 26 \\
dr corsi & 66 & free speech & 38 & deep state & 24 \\
broward county & 65 & thank q & 34 & q said & 22 \\
free speech & 64 & trust plan & 33 & john brennan & 22 \\
school shooting & 63 & white hat & 32 & president trump & 21 \\
\toprule
\end{tabular}}
\caption{\rcbts and \rga popular bigrams five days before and after quarantined.}
\label{tbl:reddit_qanon_bigrams}
\end{table}

\subsection{What are Reddit Users Discussing?}\label{sec:reddit-bigrams}

In Table~\ref{tbl:reddit_qanon_bigrams}, we list the ten most popular bigrams on \rcbts five days before it was quarantined (\emph{CBTS before quarantine} on the table); \rga five days after \rcbts was banned (\emph{GA after CBTS ban}); and \rga five days before it was quarantined (\emph{GA before quarantine}).

Discussions on \rcbts five days before it was quarantined mostly focus on QAnon topics (``q post'' mentioned $151$ times, ``deep state'' $143$ times).
We analyze the term ``false flag'' by manually inspecting all $81$ posts that include this bigram.
Users used ``false flag'' as ``false warning'' regarding the ``deep state'' passing a gun control law due to a ``school shooting'' in ``Broward county'' that week~\cite{browardshooting}. 
Manual inspection shows that most of the ``social medium'' posts refer to the same event due to users discussing the students' broadcasts of the horrors on social media.
Other ``social media'' posts discuss the ban of users. 
Lastly, we find the bigram ``dr corsi'' referring to Jerome Corsi, a conspiracy theorist, author, and academic~\cite{corsi}.
Overall, we do not detect clear signs that users knew their community would shut down.

Then, we look into the ten popular bigrams in \rga for the first five days after \rcbts gets banned. 
This time, discussions focus on users blaming the ``deep state'' for the censorship of ``social media'' by banning their ``cbts stream'' subreddit.
They refer to ``free speech'' being censored with this ``action performed.''
We manually inspect all $87$ posts that include the ``social medium'' bigram to investigate mentions of alternative platforms, finding only $1$ post suggesting alternatives:

\noindent
\blockquote{\emph{I've tried to go over to \emph{Voat} and I just don't like it [...].}}

On the other hand, users insist to ``thank q'' and ``trust the plan.''
The posts that refer to alternatives are extremely low, showing that Reddit users do not organize a migration, even after \rcbts gets banned.
During the last active days of \rga, users seem oblivious to their imminent quarantine, as all topics of discussion point to regular QAnon movement themes.

\descr{Key Findings.}
There are no signs that users suspected the ban of \rcbts, as discussions seem to be expected QAnon topics~\cite{papasavva2021qoincidence,papasavva2022gospel}, like ``deep state,'' ``q post,'' ``red pill,'' ``free speech,'' and ``red shoes''.\footnote{Reference to the use of children's skin to make shoes to hint to others that they are pedophiles~\cite{redshoes}.}
Even after \rcbts is banned, discussion or \rga do not focus on alternative platforms.

\begin{table*}[!htbp]
\centering
\small
\smallskip
\begin{tabular}{l r | l r | l r | l r}
\toprule
\multicolumn{2}{c}{Voat Q before} &  \multicolumn{2}{c}{Voat Q after} & \multicolumn{2}{|c}{Voat News before} &  \multicolumn{2}{c}{Voat News after}\\
\midrule
\textbf{Bigram} & \textbf{\emph{f}} & \textbf{Bigram} & \textbf{\emph{f}} & \textbf{Bigram} & \textbf{\emph{f}} & \textbf{Bigram} & \textbf{\emph{f}} \\
\midrule
president trump & 207  & zee bad & 214 & account deleted & 33 & stimulus bill & 5\\
martial law & 205 & bad jew & 211 & covid vaccine & 15 & still beating & 5\\
deep state & 199 & merry christmas &  195 & prc government & 13 & comment section & 5\\
supreme court & 131 & god bless & 128 & united state & 11 & covid vaccine & 4\\
united state & 131 & wwg wga & 90 & covid death & 9 & account deleted & 4\\
john robert & 127 & q post & 78 & san fransisco & 9 & election fraud & 4\\
election fraud & 120 & gonna miss & 77 & trade fraud & 7 & opinion analysis & 4\\
lin wood & 115 & greatawakening win & 71 & opinion analysis & 6 & analysis misleading & 4\\
fake news & 105 & q larp & 62 & analysis misleading & 6 & misleading title & 4\\
q post & 104 & free speech & 55 &  misleading title & 6 & open source & 4\\
\toprule
\end{tabular}
\caption{Popular bigrams on Voat QAnon and baseline subverses five days before and after the Voat shutdown announcement.}
\label{tbl:voat_qanon_bigrams}
\end{table*}

\subsection{What are Voat Users Discussing?}\label{sec:voat-bigrams}
Table~\ref{tbl:voat_qanon_bigrams} lists the ten most popular bigrams on \greatawakening, \qrv, and \vnews before (Dec 16 - Dec 20) and after (Dec 21 - Dec 25) the Voat shutdown announcement; (\emph{Voat Q before} and \emph{Voat News before}, and \emph{Voat Q after} and \emph{Voat News after}, respectively).  

The discussions before the Voat shutdown announcement focus on expected QAnon topics~\cite{papasavva2021qoincidence,papasavva2022gospel}, like ``president trump,'' ``martial law,'' ``deep state,'' ``election fraud,'' and ``q posts.''
After the announcement, the most popular bigram on QAnon subverses is ``zee bad.''
Adherents use \emph{Zee} as a short for \emph{Nazi} to describe anyone \emph{opposed} to their ideas. 
Users also refer to ``bad jew'' as they blame Jews and Nazis for taking down Voat.
Other popular bigrams are ``merry christmas,'' ``god bless,'' ``wwg wga,'' and ``gonna miss,'' which are posts sharing wishes and last goodbyes before the shutdown.

The bigram, ``q post'' appears on QAnon subverses before and after the announcement, but \emph{does not} carry the same meaning.
The ``q post'' term before the Voat shutdown announcement refers to actual Q drops, whereas later use of the term refers to the \emph{action} of Q posting on Voat or 4chan to tell users where to go.
This is further evidence of Q recommending users to migrate to \qrv after the Reddit bans.
Many users expect Q to post on Voat or other platforms to let them know where they should go: 

\noindent
\blockquote{\emph{Q will post on here!}}

\noindent
\blockquote{\emph{Wait for Q to post [...]. Q recommended QRV.}}

The term ``q larp'' stands for ``Q Live Action Role Playing.''
Previous work~\cite{papasavva2021qoincidence} explains that this term possibly demonstrates adherents losing faith in Q, explaining that Q is just role-playing.
Manual examination of our dataset aligns with these findings.
More specifically, users who are angry with Voat shutting down write: 

\noindent
\blockquote{\emph{Q-LARP is a proven jewish hoax.}}

\noindent
\blockquote{\emph{Q-LARP is entirely owned by jews.}}

\noindent
\blockquote{\emph{Q-LARP exists to create an Orwellian narrative about the bad people being arrested,}} and many more.

\noindent
Although the users agreeing to this narrative are but a handful, they created enough content to achieve the top ten most popular bigrams, trying to convince their compatriots that Q will not come to save them now. 
Lastly, ``greatawakening win'' refers to the forum \emph{greatawakening.win} (\emph{GA.win} henceforth), and ``free speech'' appears in discussions of alternatives like Poal, Parler, and Gab.

On \vnews before the announcement, users discussed COVID, the government, the US, fraud, and misleading news.
After the announcement, discussions are mostly similar, except they include ``open source'' (referring to alternative social networks) and ``still beating'' (referring to Voat being alive still).

\descr{Key Findings.}
Similarly to Reddit, Voat users mostly focus on expressing their frustration with their platform going down.
Contrary to Reddit, Voat users immediately started discussing alternatives after the shutdown announcement.
Discussions around ``q post'' changed meaning before and after the announcement, as users were \emph{waiting for Q} to post and tell them where to go, similar to what Q posted on 4chan when Reddit banned QAnon subreddits. 

\begin{table*}[!htbp]
\centering
\small
\smallskip
\begin{tabular}{l r | l r | l r | l r}
\toprule
\multicolumn{2}{c}{Poal Q before} &  \multicolumn{2}{c}{Poal Q after} & \multicolumn{2}{|c}{Poal News before} &  \multicolumn{2}{c}{Poal News after}\\
\midrule
\textbf{Bigram} & \textbf{\emph{f}} & \textbf{Bigram} & \textbf{\emph{f}} & \textbf{Bigram} & \textbf{\emph{f}} & \textbf{Bigram} & \textbf{\emph{f}} \\
\midrule
president trump & 43  & merry christmas & 111 & covid vaccine & 5 & covid death & 9\\
supreme court & 42 & president trump & 86 & robert f & 4 & merry christmas & 9\\
election fraud & 23 & election fraud & 33 & f kennedy & 4 & covid vaccine & 8\\
dominion voting & 20 & united state & 32 & kennedy jr & 4 & immune system & 8\\
covid vaccine & 15 & deep state & 26 & julian assange & 4 & news network & 8\\
white house & 15 & electoral college & 25 & credit score & 3 & deep state & 7\\
adverse reaction & 15 & insurrection act & 23 & billion dollar & 3 & common cold & 7\\
live tv & 14 & jan th & 23 & bill gate & 3 & america news & 7\\
deep state & 14 & white house & 22 & gate vaccine & 3 & white people & 6\\
insurrection act & 13 & lin wood & 20 &  gate foundation & 3 & voat shut & 3\\
\toprule
\end{tabular}
\caption{Popular bigrams on Poal QAnon and baseline subverses five days before and after the Voat shutdown announcement.}
\label{tbl:poal_qanon_bigrams}
\end{table*}

\subsection{What are Poal Users Discussing?}\label{sec:poal-bigrams}
Finally, Table~\ref{tbl:poal_qanon_bigrams} lists the most popular bigrams of the QAnon and baseline subverses on Poal, five days before the Voat shutdown announcement (between Dec 16 and Dec 20), shown as \emph{Poal Q before} and \emph{Poal News before} on the table, respectively, along with the most popular bigrams ten days after the announcement (between Dec 21 and Dec 30), shown as \emph{Poal Q after} and \emph{Poal News after}.
We extend the `after' data we use for Poal to capture discussions possibly related to the Voat shutdown and users migrating there. 

On Poal, we find expected QAnon discussions regarding Trump, election fraud, and the deep state before the Voat shutdown announcement (\emph{Poal Q before}).
We also find posts referring to ``adverse reaction'' and ``insurrection act.''
Manual inspection indicates that ``adverse reaction'' refers to COVID-19 vaccines.
The bigram ``insurrection act'' refers to posts that question whether Donald Trump should invoke the Insurrection Act to arrest the ones responsible for the election fraud: \emph{``[...]There is only ONE option left, and that is invoke the Insurrection Act [...] to get our Republic back.''}

After the Voat shutdown announcement, the most popular bigram is ``merry christmas;'' wishes from old users and newcomers alike.
Other topics include expected QAnon discussions around ``president trump,'' ``election fraud,'' ``united state,'' and ``deep state.''

Disturbingly, the term ``\emph{jan th}'' appears among the top ten popular bigrams. 
Manual inspection indicates that this is due to users discussing plans for the January 6, 2021, attack on the US Capitol.
The data used for this analysis (\emph{Poal Q after} in Table~\ref{tbl:poal_qanon_bigrams}) spans December 21, 2020, and December 30, 2020, which is more than a week before the incident.
Thus, discussions regarding the US Capitol attack shown here are \emph{premeditation} and \emph{planning} discussions.
We quote some comments regarding the planning of the January 6 US Capitol attack here, as posted by users before the attack: 

\noindent
\blockquote{\emph{
I thought, it wouldn't start until after Jan 6th, but now it is clear that things are too broken, there seems to be no point in waiting [...]}}

\noindent
\blockquote{\emph{
Trump wants a MASSIVE protest in DC on Jan 6, 2021.}}

\noindent
\blockquote{\emph{
Jan 6th D.C., Capitol Hill is a party/rally no PATRIOT is going to want to miss!}}

\noindent
\blockquote{\emph{
President Trump Calls For Protest on Jan 6., Be There, Will Be Wild}}

\noindent
\blockquote{\emph{
Jan 6th will be D-Day.}}

Regarding Poal \snews, discussions focus on COVID and the Bill Gates conspiracy~\cite{gatevaccine} as the bigrams ``billion dollar,'' ``gate vaccine,''  ``bill gate,'' and ``gate foundation'' feature prominently.
Other discussions focus on John F. Kennedy and his son, Kennedy Jr. (a common QAnon belief that Kennedy Jr. is alive and a Donald Trump ally~\cite{jfkennedy2022jr}).
Users also refer to the founder of WikiLeaks, Julian Assange.
Overall, discussions on \snews tend to be very conspiratory.
After the Voat shutdown announcement, discussions focus on COVID-19, Christmas wishes, American news, and ``white people.''
Although in small frequency, the Voat shutdown event also appears in the top ten popular bigrams on Poal \snews.  

\descr{Key Findings.}
We exposed Poal as a platform that hosts potentially dangerous individuals and controversial discussions.
Poal users, half of which are Voat migrants, discussed and organized the January 6th insurrection at the US Capitol \emph{before} it took place.
Last, \snews is far less active than the QAnon-focused subverses.

\begin{figure*}[!t]
\centering
\subfigure[Reddit]{\includegraphics[width=0.7\textwidth]{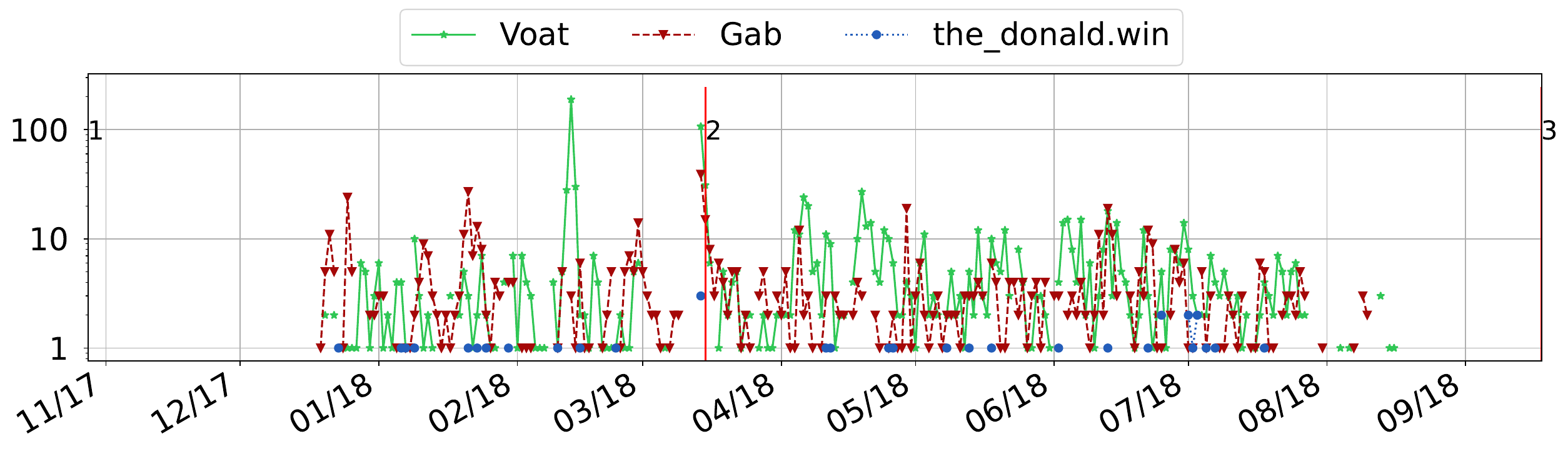}\label{fig:other_socials_on_reddit}}
\subfigure[Voat]{\includegraphics[width=0.7\textwidth]{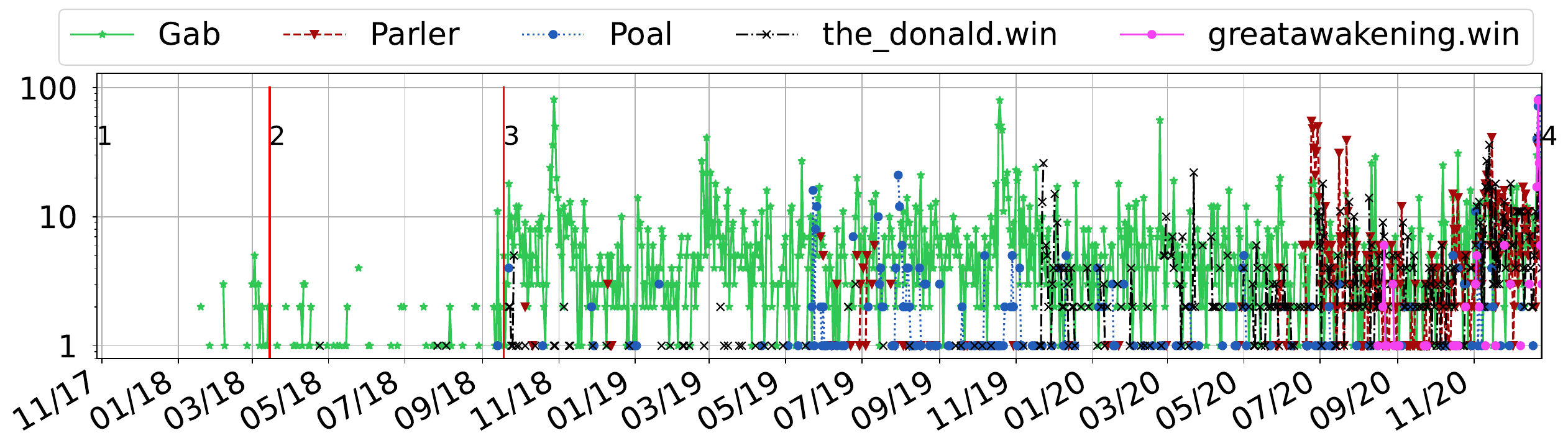}\label{fig:other_socials_on_voat}}
\caption{Alternative social networks mentioned on (a) Reddit and (b) Voat QAnon subreddits and subverses, respectively.}
\label{fig:other_socials}
\end{figure*}

\subsection{Mentions to Reddit Alternatives.}\label{sec:reddit_alternatives}
Our work focuses on understanding the differences between Reddit-to-Voat and Voat-to-Poal QAnon-focused community migration.
Hence, we dive deeper into Reddit posts 
to investigate whether Reddit users mentioned any Reddit alternatives and when such discussions, if any, took place.
To do so, we plot the daily number of mentions to Voat, Gab, and thedonald.win over the lifetime of \rcbts and \rga on Reddit in Figure~\ref{fig:other_socials_on_reddit}.
We look specifically for these social networks as they are the ones we came across very often when we manually inspected our dataset for the analysis discussed in Section~\ref{sec:reddit-bigrams}, above.

In Figure~\ref{fig:other_socials_on_reddit}, we observe that Voat was the social network most discussed on Reddit, with over $1.2K$ mentions.
Gab comes second with $708$ mentions, and the\_donald only has $38$.
Importantly, we observe two peaks in Voat mentions on Reddit: 13 February 2018 with $187$ mentions and 14 March 2018 with $102$ mentions.
To understand the discussions around Voat on the first peak, we select and read all these posts to qualitatively discuss the patterns and discussions of these posts.
We find that about half of these posts express the frustration of the \rcbts users towards a moderator who deleted many posts.
We list some of these posts:

\noindent
\blockquote{\emph{
100+ posts are being REMOVED EVERY DAY! Our new site is voat.co}}

\noindent
\blockquote{\emph{
After a year of subverting voat [mod user] has finally been outed}}

\noindent
\blockquote{\emph{[...] FUCK YOU [mod user]!}}

We speculate that \rcbts moderators started deleting posts attempting to abide by Reddit policies.
It is clear that \rcbts users were unhappy with this, as many suggested Voat as an alternative.
The second peak on Voat mentions shows up on the day that \rcbts was actually banned, with the majority being a coordinated action by various users:

\noindent
\blockquote{\emph{Let's go to voat!! Fuck reddit!}} 

\noindent
\blockquote{\emph{Post this link everywhere: voat.co/v/greatawakening}}

\subsection{Mentions to Voat Alternatives.}
We plot the daily number of mentions to Poal, Gab, Parler, thedonald.win, and GA.win over the last six months of Voat in Figure~\ref{fig:other_socials}.
We selected these five social networks as they are the ones that appear most frequently when we manually inspect our dataset for the analysis discussed in Section~\ref{sec:voat-bigrams}.

Overall, Gab was mentioned $5.2K$ times, Parler was mentioned $1.3K$ times, thedonald.win $1.2K$ times, Poal was mentioned $621$ times, and GA.win $181$ times.
Although Gab is the most popular social network mentioned, it rarely appears during the final days of Voat.
Poal is the most popular platform in comments during the last week of Voat, with $79$ mentions on December 24, 2020, alone.

To better understand the discussions around Poal, we collect the posts that refer to Poal during the last four days of Voat and find the most popular bigrams:
``open source,'' ``destroy voat,'' ``voat try,'' ``soap box,'' ``source decentralized,'' ``blog diasporafoundation,'' ``buddypress category,'' ``user signin,'' ``try friend,'' ``free speech,'' and ``go poal.''
Although many users believe Poal was created to ``destroy voat,'' many others support and urge others to move there (``go poal,'' ``try friend''). 
More specifically, many users mention that they already created an account on Poal and explain how the ``user signin'' process works, how Poal supports ``free speech,'' etc.
In addition, we manually inspect these posts, finding many posts encouraging users to join Poal:

\noindent\blockquote{\emph{Wanted to encourage everyone to head to poal.}}

\noindent
\blockquote{\emph{HEY GUYS LET'S ALL MIGRATE TO POAL!}}

\noindent
\blockquote{\emph{POAL the Place to Go!}}

\section{Sentiment Analysis}\label{sec:sentiment}
Taking note of how many Reddit users urged their fellow QAnon adherents to follow them on Voat, and from how Voat users urged their fellow ``patriots'' to join them on Poal (or other alternatives), we perform sentiment analysis to understand the sentiment of Reddit and Voat users towards the suggested alternative platforms.
We use sentiment analysis for this analysis since users discussing alternatives is a similar concept to users rating products, on which sentiment detection performs well~\cite{kumar2019fusion}.

To this end, we employ a pre-trained VADER sentiment detection model~\cite{hutto2014vader}. 
The model takes as input a string and responds with four values.
\emph{neg} indicates negative sentiment and ranges between $0$ to $+1$. 
VADER returns the same range for \emph{pos}, for positive sentiment, and \emph{neut}, which indicates neutrality.
The model also returns \emph{comp}, which stands for \emph{compound}: a normalization between negative and positive sentiment.

\begin{figure}
\centering
\subfigure[Reddit]{\includegraphics[width=0.62\columnwidth]{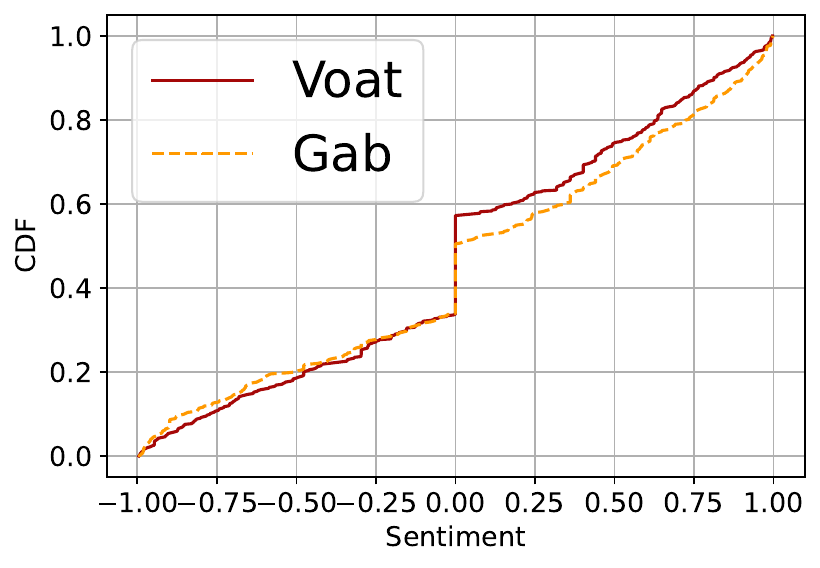}\label{fig:cdf_reddit}}
\subfigure[Voat]{\includegraphics[width=0.62\columnwidth]{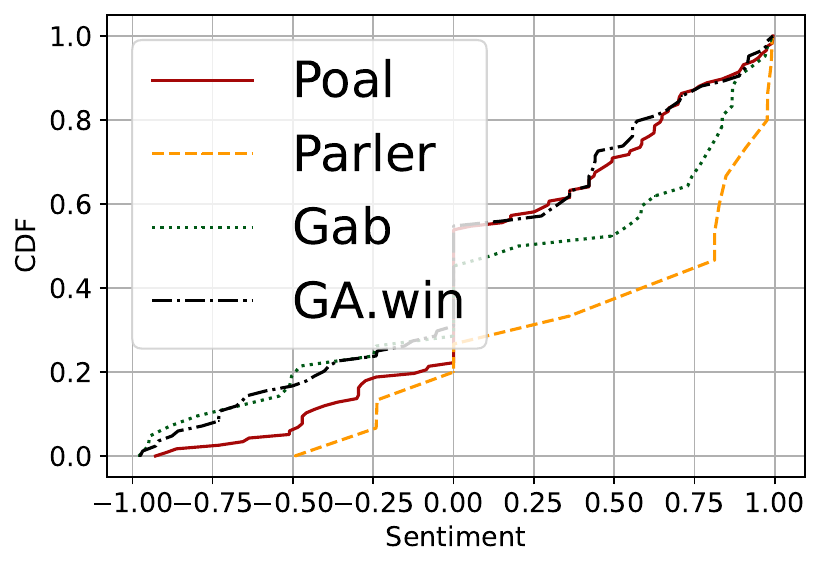}\label{fig:cdf_voat}}
\caption{CDF plots of the compound score on (a) Reddit and (b) on Voat discussions mentioning alternatives.}
\label{fig:CDFsentiment_both}
\end{figure}

For Reddit, we analyzed posts that mentioned Voat and Gab. 
We do not restrict the analysis to specific periods, as references to alternatives on Reddit mainly emerged after \rcbts moderators began removing content.
We omit sentiment analysis of the\_donald mentions as it was only mentioned $38$ times, usually to refer users to a discussion there, and not to suggest it as an alternative.
We filtered out posts with no sentiment, such as those that only contained links to Voat.
This results in $546$ posts mentioning Voat, and $405$  mentioning Gab.
We feed these posts to VADER and use the \emph{comp} score to plot the Cumulative Distribution Function (CDF) in Figure~\ref{fig:cdf_reddit}.
Overall, about $42.7\%$ ($233$ posts) of all the posts that mention Voat have positive sentiment polarity, with $25.5\%$ ($139$ posts) of them having strong positive sentiment polarity towards Voat (\emph{comp >} $+0.5$).
$33.7\%$ ($184$ posts) of the posts mentioning Voat seem to have negative sentiment polarity (\emph{comp < }$0$).

Regarding posts mentioning Gab, $49.4\%$ ($200$ posts) show positive polarity, with $30\%$ ($125$ posts) having strong positive polarity and $34\%$ of the posts ($138$) depict a negative sentiment polarity.
We use the Kolmogorov-Smirnov test~\cite{massey1951kolmogorov} to detect whether the distributions are statistically significant and do not reject the null hypothesis for the distributions between Gab and Voat, which means that the two distributions are not significantly different ($p = 0.23$).

For Voat, we analyze posts that mention Poal, Parler, Gab, and GA.win after the Voat shutdown announcement.
We focus on these social networks due to their popularity when we manually inspect our QAnon community dataset during the last days of Voat.
Following the same parsing mentioned above, we end up with $118$ posts that mention Poal, $17$ posts for Parler, $49$ for Gab, and $85$ for GA.win that are used to plot the CDF of the compound value in Figure~\ref{fig:cdf_voat}.
Overall, $45.7\%$ of the posts ($54$) that mention Poal have positive sentiment polarity, with $28.8\%$ of them ($34$) indicating strong positive sentiment.
Only $\sim8\%$ of the posts ($10$) indicate a strong negative sentiment towards Poal (\emph{comp <} $-0.5$).
Regarding Parler, $58.8\%$ of the posts indicate strong positive sentiment polarity ($10$ posts), and $17.6\%$ of them exhibit negative polarity ($3$ posts).
$40.8\%$ of the posts ($20$) that mention Gab have strong positive sentiment polarity, and $24.5\%$ of them ($12$) exhibit a negative one.
Lastly, $27\%$ of the posts on Voat that mention GA.win achieve strong positive sentiment polarity ($23$), and $30.5\%$ of them ($26$) have a negative one.

Although Parler has the highest percentage of posts that indicate positive sentiment, Poal was mentioned the most during Voat's final days.
KS test rejects the null hypothesis for the distributions between Poal and Parler, Poal and Gab, Gab and GA.win, and Parler and GA.win.
That means that the four distributions are significantly different; they have no similarity ($p = <0.001$).
The distributions of Poal and GA.win are not significantly different ($p = 0.5$).
The same is true for the distributions of Parler and Gab ($p = 0.18$).

\descr{Key findings.}
On Reddit, posts mentioning Voat received both positive and negative sentiment, with Gab showing both the highest positive sentiment and the highest negative sentiment.
The two distributions are not significantly different, which indicates that users liked Gab as an alternative as much as they preferred Voat.
On the other hand, Voat was mentioned \emph{more often}, prompting users to join Voat.
After the Voat shutdown announcement, Poal was the most mentioned alternative, exhibiting as much favor as GA.win.

\section{Discussion and Conclusion}\label{sec:discussion}
This study presents a comprehensive analysis of the shutdown crisis in QAnon-focused subverses on Voat, using a mixed-methods approach, focusing on Reddit-to-Voat and Voat-to-Poal migrations. 
We now delve deeper into our findings, emphasizing the significance of studying fringe platforms like Poal, highlighting the differences between the Reddit-to-Voat and Voat-to-Poal migrations, and exploring the implications of deplatforming. 
Furthermore, we discuss potential future research directions and acknowledge the limitations of our study before concluding our work.

\descr{Why study Poal?}
As the first study to investigate Poal, we raise questions about its status as a fringe platform and its relevance to the research community. 
Considering the contentious nature of the QAnon community and the presence of users discussing controversial topics on Poal, we argue that Poal has emerged as the new Voat, characterized by echo chambers and potentially harmful communities. 
Our research provides evidence that users on Poal were involved in organizing the January 6th insurrection at the US Capitol prior to the event, highlighting the platform's association with dangerous individuals. 

Poal represents just one of many fringe communities where discussions of a controversial nature occur, often with lower activity compared to platforms like Reddit. 
However, delving deeper into these communities is crucial for a comprehensive understanding of their content and implications, leading us to the next point of discussion.

\descr{How do the two migrations differ?}
The Reddit-to-Voat migration seems less organized than the Voat-to-Poal one. 
This disparity can be attributed to the fact that Voat users were aware of the impending shutdown of their platform, providing them with time to plan.  
The Reddit bans triggered some discussions about alternative platforms, but activity on Voat did not peak following the first ban of \rcbts, as users migrated to another QAnon subreddit: \rga. 
Given the opportunity, users prefer to join a different community within the same platform that aligns with their beliefs, rather than migrating to an entirely new platform, aligned with the findings of Engel et al.~\cite{engel2022characterizing}.
The activity on Voat notably increased only after the last \emph{active} QAnon subreddit was banned. 
Furthermore, we found evidence indicating that Q, the ``mastermind'' behind QAnon, recommended users to migrate to \qrv after the ban of \rga.

Indeed, our analysis revealed a noteworthy phenomenon regarding the migration of Reddit users to Voat and the subsequent emergence of \qrv as we observe a sudden influx of users to \qrv immediately after the posts from Q urging users to move there. 
We note that the posts from Q took place on 4chan, which resulted in activity on Voat, proceeding Reddit's bans, demonstrating how well-organized and multi-platform-based the conspiracy is. 
Gab and Voat were the most popular preferred alternatives to Reddit, but both platforms exhibited similar patterns of sentiment polarity, with no significant difference in sentiment distribution. 
This suggests that Reddit users did not reach a clear consensus on a preferred alternative platform, adding to the complexity of the migration dynamics. 

On the contrary, upon the announcement of the impending shutdown of Voat, users on the platform quickly engaged in discussions about alternative platforms, with Poal emerging as the clear preference for Voat users. 
This finding highlights the organized nature of a premeditated migration, with a larger percentage of users moving to a specific platform. 

\descr{What are the implications of these bans?}
Deplatforming garnered attention from researchers in 2017, about two years after the initial Reddit ban. 
Banned communities either leave the platform or reduce hate speech~\cite{chandrasekharan2017you}; however, other research~\cite{papasavva2021qoincidence, ribeiro2021migrations, jhaver2021evaluating, dehghan2022politicization} contends that displaced users often migrate to more permissive platforms, perpetuating harmful discussions, aligned with our findings.
Voat's pro-QAnon users migrated to Poal, where discussions veered into terrorism and potentially harmful actions, e.g., the US Capitol raid. 
Deplatforming presents challenges for computational social scientists in tracking diverse communities across platforms.
We recognize that mainstream platforms must remove controversial content in an attempt to ensure a safe environment. 
However, our findings reveal unmoderated fringe communities like Poal continue to host harmful activities, complicating the battle against controversial content. 
Assessing the long-term impact of deplatforming is complex due to evolving online ecosystems. 
Continued research is necessary to comprehend the effectiveness of deplatforming strategies and their consequences. 
Our study offers insights into deliberate and coordinated online migrations, contributing to this ongoing research.

\descr{Directions for future work.}
While our current study does not delve into toxicity comparisons between Poal and Voat, this is an area of potential research. 
Additionally, extending our investigation to include data from platforms like Parler, Gab, and GA.win could provide insights into the QAnon community's responses to significant events, such as Q's re-emergence after years of silence. 
Finally, examining the attraction of dangerous individuals to Poal due to its relaxed moderation policies warrants further exploration.

\descr{Limitations.}
\emph{Datasets}:
During the early stages, we faced challenges collecting data from greatawakening.win and Gab. 
The technical difficulties, such as greatawakening.win's random submissions and limited data accessibility, made it difficult to ensure the completeness of our dataset, hence we had to omit these platforms.

\noindent
\emph{Platform Selection}:
Due to their operational and aesthetic similarities, we focused solely on Reddit, Voat, and Poal, leaving out other platforms like Gab, Parler, and greatawakening.win. 
While our findings show interest in these platforms, future studies could provide additional insights into user displacement dynamics.

\noindent
\emph{Username Changes}:
In Section~\ref{subsec:useractivity}, we only considered identical usernames across platforms. 
This approach has limitations, as users may change their usernames when migrating or ``squat'' someone else's username to impersonate them~\cite{mariconti2016allowing,mariconti2017whats}.
Despite this limitation, examining identical usernames offered valuable insights into migration patterns.

\descr{Conclusion.}
This study provides valuable insights into the migration behaviors of QAnon adherents across Reddit, Voat, and Poal. 
Notable findings include the organized migration patterns from Voat to Poal, highlighting the potential benefits of advance notice regarding platform shutdowns. 
Additionally, our study highlights the significance of investigating fringe platforms like Poal, where dangerous individuals congregate, and discussions about the January 6th, 2021 attack on the US Capitol transpire. 
Furthermore, the challenges posed by deplatforming efforts are evident, as these communities disperse to alternative platforms with laxer moderation, complicating the detection and mitigation of harmful content. 
This research contributes to a deeper understanding of online migration dynamics and emphasizes the need for further investigation in the ever-evolving landscape of online communities.

\begin{acks}
This project was funded by the UK EPSRC grant EP/S022503/1, which supports the UCL Center for Doctoral Training in Cybersecurity.
The lead author extends sincere thanks to Arianna Trozze for her valuable feedback and meticulous editorial assistance.
\end{acks}

\bibliographystyle{ACM-Reference-Format}
\bibliography{references}

\end{document}